\documentclass[twocolumn,showpacs,preprintnumbers,amsmath,amssymb,nofootinbib]{revtex4}
\usepackage{dcolumn}
\usepackage{bm}
\usepackage{tabularx}
\usepackage{array}
\usepackage{graphics}
\usepackage{graphicx}
\usepackage{psfrag}
\usepackage{epsfig}
\usepackage{amsmath}
\usepackage{amssymb}
\usepackage{citesort}
\usepackage{verbatim}
\usepackage{color}

\newcommand{\bea}{\begin{eqnarray}}
\newcommand{\eea}{\end{eqnarray}}
\newcommand{\beq}{\begin{equation}}
\newcommand{\eeq}{\end{equation}}
\newcommand{\gev}{{\rm GeV}}
\newcommand{\nn}{\nonumber}

\newcommand{\pdir}{p\kern -5.2pt\raise 0.2ex\hbox {/}}
\newcommand{\vdir}{v\kern -5.75pt\raise 0.15ex\hbox {/}}
\newcommand{\kdir}{k\kern -5.75pt\raise 0.15ex\hbox {/}}
\newcommand{\epsdir}{\epsilon\kern -5.0pt\raise 0.15ex\hbox {/}}
\newcommand{\bvdir}{\bar{v}\kern -5.75pt\raise 0.15ex\hbox {/}}
\newcommand{\Ddir}{D\kern -7.75pt\raise 0.20ex\hbox {/}}
\newcommand{\ldir}{l\kern -5.0pt\raise 0.2ex\hbox{/}}
\newcommand{\varepsdir}{\varepsilon\kern -5.5pt\raise 0.15ex\hbox{/}}

\newcommand{\mawi}{m_{\scriptstyle{ \rm AWI}}^{(0)} }

\def\Tr{\operatorname{Tr}}

\def\negcdot{\negmedspace\cdot\negmedspace}

\begin{document}

\preprint{LPT Orsay 09-34}
\vspace*{22mm}
\title{$g_{B^\ast B\pi}$-coupling in the static heavy quark limit}

\author{Damir Be\'cirevi\'c}
 \email{Damir.Becirevic@th.u-psud.fr}
\affiliation{%
Laboratoire de Physique Th\'eorique (B\^at 210), Universit\'e Paris Sud,  
Centre d'Orsay, 91405 Orsay-Cedex, France}

\author{Benoit Blossier} 
\email{Benoit.Blossier@th.u-psud.fr}
\affiliation{%
Laboratoire de Physique Th\'eorique (B\^at 210), Universit\'e Paris Sud,  
Centre d'Orsay, 91405 Orsay-Cedex, France}

\author{Emmanuel Chang}
 \email{Emmanuel.Chang@th.u-psud.fr}
\affiliation{%
Laboratoire de Physique Th\'eorique (B\^at 210), Universit\'e Paris Sud,  
Centre d'Orsay, 91405 Orsay-Cedex, France}

\author{Benjamin Haas}
 \email{Benjamin.Haas@th.u-psud.fr}
\affiliation{%
Laboratoire de Physique Th\'eorique (B\^at 210), Universit\'e Paris Sud,  
Centre d'Orsay, 91405 Orsay-Cedex, France}

\date{\today}

\begin{abstract}
By means of QCD simulations on the lattice, we compute the coupling of the  heavy-light mesons to a soft pion in the static heavy quark limit. 
The gauge field configurations used in this calculations include the effect of $N_f=2$ dynamical Wilson quarks, while for the static quark propagator 
we use its improved form (so called HYP).  On the basis of our results we obtain that the coupling $ \hat g  =0.44\pm 0.03{}^{+0.07}_{-0.00}$, where the second error is flat (not gaussian). 
\end{abstract}

\maketitle

\section{\label{Introduction}Introduction}
The static quark limit of QCD offers a simplified framework to solving the non-perturbative dynamics of light degrees of freedom in the heavy-light systems. 
That dynamics is constrained by heavy quark symmetry (HQS): it is blind to the heavy quark flavor and its spin. As a result the total angular momentum of the light 
degrees of freedom becomes a good quantum number ($j_\ell^P$), and therefore the physical heavy-light mesons come in mass-degenerate doublets. 
In phenomenological applications the most interesting information involves the lowest lying doublet, the one with $j_\ell^P=(1/2)^-$,  consisting of a pseudoscalar and a vector meson, 
such as ($B_q$,$B^\ast_q$) or ($D_q$,$D^\ast_q$) states, where $q\in \{u,d,s\}$. When studying any phenomenologically 
interesting  quantity  from the QCD simulations on the  lattice that includes heavy-light mesons (decay constants, various form factors, bag parameters and so on), one of the major 
sources of systematic uncertainty is related to the necessity to make chiral extrapolations. The reason is that the physical light quarks, which are expected to most significantly modify 
the structure of the QCD vacuum,  are much lighter than the ones that are directly simulated on the lattice, $m_q\gg m_{u,d}$.  Here by ``q" we label the light quark masses that are attainable from the lattice. Since the QCD dynamics with very light quarks 
is bound to be strongly affected by the effects of spontaneous chiral symmetry breaking, a more suitable (theoretically more controllable) way to guide such extrapolations is by using the expressions 
derived in  heavy meson chiral perturbation theory (HMChPT), which is an effective theory built on the combination of HQS and 
the spontaneous chiral symmetry breaking  [$SU(N_f)_L\otimes SU(N_f)_R \to SU(N_f)_V$].  Its Lagrangian is given by~\cite{Casalbuoni}
\bea \label{heavy}
&&\hspace*{-5mm}{\cal L}_{\rm heavy}=-{\rm tr}_a \Tr[\overline{H}_a i v \negcdot D_{ba}H_b]+ \hat g\ {\rm tr}_a\Tr[\overline{H}_aH_b \gamma_\mu 
{\bf A}_{ba}^\mu \gamma_5]\,,\cr
&&\hspace*{-5mm}D^\mu_{ba}H_b =  \partial^\mu H_a -  H_b {1\over 2}[ \xi^\dagger \partial_\mu \xi + 
\xi \partial_\mu \xi^\dagger ]_{ba}\;,\cr
&&\hspace*{-3mm}{\bf A}_\mu^{ab}
= {i\over 2}[ \xi^\dagger \partial_\mu \xi - 
\xi \partial_\mu \xi^\dagger ]_{ab} \;,
\eea
where
\bea
H_a(v) = {1 +  \vdir \over 2} \left[ P^{\ast\ a}_\mu (v)\gamma_\mu - P^a (v)\gamma_5
\right]\;,\eea
is the heavy meson doublet field containing the pseudoscalar, $P^a (v)$, and the vector meson field, $P^{\ast \ a} (v)$. In the above formulas, the indices $a,b$ run over the light quark flavors, 
$\xi =  \exp\left(i \Phi/f \right)$, with $\Phi$ being the matrix of $(N_f^2-1)$ pseudo-Goldstone bosons, and ``$f$" is the pion decay constant in the chiral limit. 
We see that the term connecting the Goldstone boson  (${\bf A}_\mu$) with  the heavy-meson doublet [$H(v)$] is proportional to the coupling $\hat g$, which will therefore enter into every expression 
related to physics of heavy-light mesons  with $j_\ell^P=(1/2)^-$ when the  chiral loop corrections are included.~\footnote{A special attention should be given to the problem related to the presence of the nearby excited states as discussed in ref.~\cite{su2}. 
Any precision lattice calculation cannot be fully trusted if the chiral extrapolations are made without discussing the problem of discerning the mixing with the $j_\ell^P=(1/2)^+$ states in the chiral loop diagrams.} Being the parameter of effective theory, its value cannot be predicted but should be fixed in some other way. 
It can be related to  the measured decay width $\Gamma(D^\ast \to D\pi)$~\cite{CLEO}, with the resulting value $\hat g_{\rm charm}=0.61(7)$. That value turned out to be much larger than predicted 
by all of the QCD sum rule approaches~\cite{KRWY}, but consistent with some model predictions such as the one in ref.~\cite{dirac},  in which  a more detailed 
list of predictions with their references can be found. The large value for $g_{D^\ast D\pi}$-coupling was confirmed by the quenched lattice QCD study in ref.~\cite{gDDpi}, and recently also in the unquenched case~\cite{gDDpi-unq}.  
Since the charm quark is not very heavy, the use of $g_{D^\ast D\pi}$ to fix the value of $\hat g$-coupling, via
\bea
\hat g ={g_{D^\ast D\pi}\over 2 \sqrt{m_D m_{D^\ast}} } f_\pi\,,
\eea
 and its use in  chiral extrapolations of the quantities relevant to $B$-physics phenomenology may 
be dangerous mainly because of the potentially large ${\cal O}(1/m_c^n)$-corrections. 
Unfortunately the decay $B^\ast \to B\pi$ is kinematically forbidden and therefore, to determine the size of $\hat g$, we have to resort to a non-perturbative 
approach to QCD.  Unlike for the computation of the heavy-to-light form factors, QCD sum rules proved to be inadequate when computing $g_{D^\ast D\pi}$, most likely because 
of the use of double dispersion relations when  the radial excitations should be explicitly included in the analysis, as claimed in ref.~\cite{us9}.
In this paper, instead, we compute the $\hat g$-coupling on the lattice by using the unquenched gauge field configurations, with $N_f=2$ dynamical light quarks, and in the static heavy quark limit. 
The attempts to compute this coupling in this limit were made in ref.~\cite{g-ukqcd-orsay}, and very recently in ref.~\cite{onogi}.  On the basis of the currently available information, the coupling  $\hat g$ 
in the static limit is indeed smaller than the one obtained in the charmed heavy quark case.

In the remainder of this letter we will briefly describe the standard strategy to compute this coupling, list the correlation 
functions that are being computed to extract the bare coupling $\hat g_q$, as well as the axial vector renormalization constants. We then give details concerning the 
gauge field configurations used in this work, and present our results.

\section{Definitions and Correlation functions to be computed}

In the limit in which the heavy quark is infinitely heavy and the light quarks massless, 
the axial coupling of the charged pion to the lowest lying doublet of heavy-light mesons, $\hat g$,  is defined via~\cite{g-ukqcd-orsay}
\bea
\langle B\vert \vec  A\vert B^\ast(\varepsilon) \rangle = \hat g\ \vec \varepsilon_\lambda \,,
\eea
where the non-relativistic normalisation of states $\vert B^{(\ast)}\rangle$ is assumed,  
$  \langle B_a(v)\vert B_b(v^\prime)\rangle = \delta_{ab}\delta(v - v^\prime)$. For the heavy-light hadrons  at rest 
($\vec v=\vec v^\prime =\vec 0$),  the soft pion that couples to the axial current,  
${A}_\mu = \bar u\gamma_\mu \gamma_5 d$, is at rest too,  $|\vec q |= 0$. $\varepsilon_\mu^\lambda$ is 
the polarisation of the vector static-light meson. In the typical situation on the lattice we are away from the chiral limit ($\hat g\to \hat g_q$), and the coupling $\hat g_q$ becomes 
the axial form factor whose value should be extrapolated to the chiral limit, in which the soft pion theorem relating the matrix element of the axial current to the pionic coupling applies~\cite{g-ukqcd-orsay}.

The standard strategy to compute the above matrix element on the lattice consists in evaluating the following  correlation functions:
\bea\label{eq:correlators}
C_2(t) &=&\langle\sum_{\vec x} P(x) P^\dagger(0)\rangle_{_U} \stackrel{^{\rm HQS}}{=}\frac{1}{3}\langle\sum_{i,\vec x} V_i(x) V_i^\dagger(0)\rangle_{_U}\cr
& =&
\langle\sum_{\vec x} {\rm Tr}\left[ {1+\gamma_0\over 2}W_x^0 \gamma_5 {\cal S}_{u,d}(0,x) \gamma_5\right]\rangle_{_U} , 
\\
C_3(t_y,t_x)&=&\langle\sum_{i,\vec x,\vec y} V_i(y) {A}_i(x) P^\dagger(0)\rangle_{_U} \cr
&=&
\langle\sum_{\vec x,\vec y} {\rm Tr}\left[ {1+\gamma_0\over 2}W_0^y \gamma_i {\cal S}_u(y,x) \gamma_i\gamma_5
{\cal S}_d(x,0)\gamma_5\right]\rangle_{_U}  ,\nn
\eea
where $\langle\dots\rangle_{_U}$ denotes the average over independent gauge field configurations, the interpolating fields are
$P= \bar h \gamma_5 q$, $V_i= \bar h \gamma_i q$, with $h(x)$ and $q(x)$ the static heavy and the light quark field, respectively. 
In what follows, we drop the dependence on $t_y$. In practice its value is fixed to one or several values as it will be specified in the text. 
In eq.~(\ref{eq:correlators}) we also expressed the correlation functions in terms of quark propagators: the light ones, ${\cal S}_{q}(x,y)$, and the static heavy one, which becomes a Wilson line, 
\bea\label{Pline}
W_x^y = \delta(\vec x-\vec y) \prod_{\tau=t_y}^{t_x-1}U^{\rm impr.}_0(\tau, \vec x)\,.
\eea
The latter is merely obtained from the discretized static heavy quark action~\cite{Eichten-Hill}
\bea \label{hqet-lagr}
{\cal L}_{\rm HQET}=\sum_x  h^\dagger(x)\left[ h(x) - U^{\rm impr.}_0(x-\hat 0)^\dagger h(x-\hat 0)\right] \,,
\eea
where for $U^{\rm impr.}_0$, the time component of the link variable, we use its improved form, obtained after applying 
the hyper-cubic blocking procedure on the original link variable,  with the parameters optimized in a way described in ref.~\cite{knechtli}, namely with $\vec \alpha =(0.75,\ 0.6,\ 0.3)$.  
That step is essential as it ensures the exponential improvement of the signal to noise ratio in the correlation functions with respect to 
what is obtained by using the simple product of link variables~\cite{alpha}.

\begin{figure}
\hspace*{-.7cm}\includegraphics[width=8.55cm,clip]{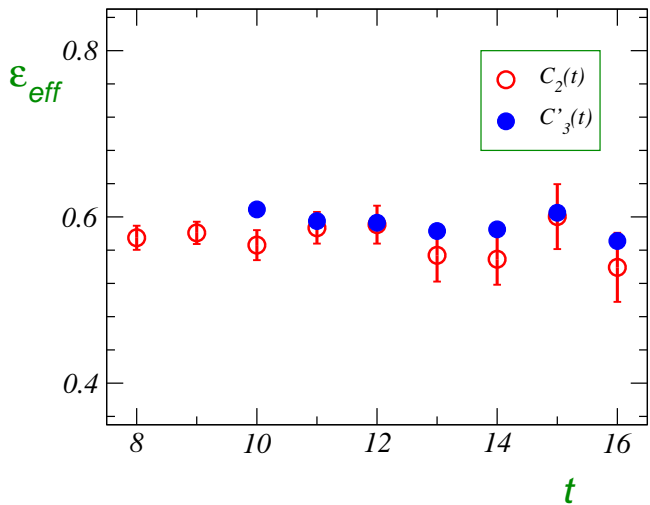}
\caption{\label{fig:3} \sl Comparison of the effective binding energy extracted from $C_2(t)$ and from $C_3'(t_x)$ for the Set 2 and $\kappa_1$ (c.f. table~\ref{tab:1}). 
Notice that the separation of the sources $t=t_x - t_{\rm fix}$ and in our case $t_{\rm fix}=-5$. }
\end{figure}

The spectral decomposition of the three point function, given in eq.~(\ref{eq:correlators}), reads
\bea
C_3(t_x) = \sum_{m,n} \Big[ {\cal Z}_n e^{-{\cal E}^q_{n} t_y} \langle B_n\vert A_i\vert B^\ast_m\rangle 
e^{-({\cal E}^q_m -{\cal E}^q_n)t_x} {\cal Z}_m  \varepsilon_i^{(m)} \Big]\,,\nn
\eea 
where the sum includes not only the ground states ($m=n=0$) but also their radial excitations 
($m,n>0$), which are heavier and thus exponentially suppressed. Note a shorthand notation,  
${\cal Z}_n = \vert \langle 0| h^\dagger \gamma_5 q| B_q\rangle\vert$, and the fact that we do not distinguish ${\cal Z}_n$ from couplings to the vector interpolating 
operator because of the HQS. 
If the non-diagonal terms in the above sum  were important  ($n\neq m$) the correlation function $C_3(t_x)$ would exhibit 
some exponential dependence in $t_x$. 
\begin{figure}
\begin{center}
\epsfig{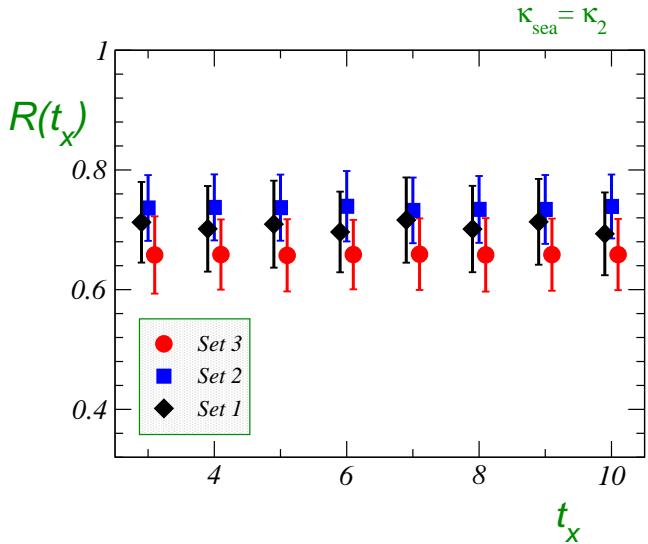}
\caption{ \label{fig:plateaus} \sl Ratio $R(t_x)$ in eq.~(\ref{gfit}) as obtained from our data for all three sets and for which $\kappa_{\rm val}=\kappa_{\rm sea}=\kappa_2$, 
with $\kappa_2$  specified in table~\ref{tab:1}. This plot also shows the flatness of the signal of  $C_3(t_x)$ defined in eq.~(\ref{eq:correlators}).
For completeness, we also note that $t_y=13$.}
\end{center}
\end{figure}
In practice, it appears that the correlation functions $C_3(t_x)$, as defined in  eq.~(\ref{eq:correlators}) are very flat ($t_x$-independent) 
for all the data sets that we use in this work and the details of which will be given in the next section (c.f. fig.~\ref{fig:plateaus}). This observation  in fact  agrees with what one can deduce from various 
quark models, and in particular from the one in ref.~\cite{dirac}.   We will therefore discard the non-diagonal terms in the spectral decomposition of  $C_3(t_x)$.
We are still left with the  problem of contamination of the desired signal ($n=0$) by the axial transitions among radial excitations, $n=m >0$. 
To solve that problem we should employ some smearing procedure and suppress the couplings of the source operators to the radial excitations.
To that purpose we use the smearing technique proposed in ref.~\cite{Boyle}, which essentially means that  --in eq.~(\ref{eq:correlators})-- the interpolating fields are replaced 
by $\bar h(x) \gamma_5 q(x)\to \bar h(x) \gamma_5 q^S(x)$,  and similarly for the source of the heavy-light vector mesons, where
\bea \label{Smearing}
q^S(x)= \sum\limits_{r=0}^{R_{\rm max}} \varphi(r)
 \sum\limits_{k=x,y,z} \left[
q({\scriptstyle x+r\hat{k}})\times \prod\limits_{i=1}^r U_k({\scriptstyle
x+(i-1)\hat{k}})\right. \cr
\left. + q({\scriptstyle x-r \hat{k}})\times
\prod\limits_{i=1}^r U_k^{\dagger}({\scriptstyle
x-i\hat{k}})\right]  \,,
\eea 
and $\varphi(r)=  e^{-r/R}(r+1/2)^2$. The link variables on the right hand side of eq.~(\ref{Smearing}) are fuzzed as discussed in ref.~\cite{g-ukqcd-orsay}. After several trials we chose the smearing 
parameters to be $R=1.3$ and $R_{\rm max}=4$, to highly enhance the overlap with ground states.  From the fits of our two-point functions computed with both the local (``loc.") and smeared sources (``sm.") to two exponentials 
on the large interval $4 \leq t\leq 15$,  we obtain  that  ${\cal Z}_0^{\rm sm.}/{\cal Z}_0^{\rm loc.}\gtrsim 45$, while $Z_1^{\rm sm.}/Z_1^{\rm loc.} < 0.05$.  More importantly, 
 ${\cal Z}_1^{\rm sm.}/{\cal Z}_0^{\rm sm.}< 0.04$, or it cannot be fitted, when it is completely absent.  We therefore deduce that our smearing is efficient and the contribution of the radial excitations 
 is most probably negligible.  To further check this point we reorganized the operators in  $C_3(t_x)$ and  fixed the transition operator [$A_i$ in eq.~(\ref{eq:correlators})] at $t=0$, one source 
 operator at $t_y\equiv t_{\rm fix} = -5$, and have let the other source operator free (c.f. also ref.~\cite{emmanuel}). 
In that situation the spectral decomposition looks as follows,
\bea\label{eqC3bis}
C_3^\prime (t_x) \simeq \sum_{n}  {\cal Z}_n^2 e^{- {\cal E}^q_{n}\ (t_x - t_{\rm fix}) } \langle B_n\vert A_i\vert B^\ast_n\rangle 
   \varepsilon_i^{(n)} \,,
\eea 
which allows us to  check whether or not its  effective binding energy, with the smeared source operators,
\bea
{\cal E}^q_{\rm eff}(t_x) = \log\left( { C_3^\prime (t_x) \over C_3^\prime (t_x+1)}\right) \,,
\eea
agrees with what is obtained from the two-point correlation functions, 
\bea
{\cal E}^q_{\rm eff}(t) = \log\left( { C_2 (t) \over C_2 (t+1)}\right) \,.
\eea
This is illustrated in fig.~\ref{fig:3}, which we find satisfactory. After these checks, we  extract $\hat g_q$ from the fit to a constant of the ratio,~\footnote{The use of index ``$q$" in $\hat g_q$ should not be confusing to the reader. Here it simply 
labels the light quark directly accessed from our lattices.} 
\bea\label{gfit}
R(t_x)={1\over 3}{C_3(t_x)\over (Z_0^{\rm sm.})^2 e^{-{\cal E}^q_{0}t_y}}\longrightarrow \hat g_q\,.
\eea
All our fits are made on the common interval, $5 \leq t_x\leq 8$.  On one ensemble of our gauge-field configurations we also checked 
that  the value of $\hat g_q$  extracted from eq.~(\ref{gfit}) is fully consistent with what is obtained if the computation is organized as in eq.~(\ref{eqC3bis}).

 \begin{table}[t]
\begin{center}
\begin{tabular}{|c|cccc|}\hline
{\phantom{\Large{l}}} \raisebox{-.1cm} {\phantom{\Large{j}}}
\hspace*{-3.5mm} Action &$ \beta$ [$a$ (fm)] & \# meas. & $\kappa_{\rm sea}$ & ref. \\ \hline\hline
{\phantom{\Large{l}}}  {\phantom{\Large{j}}}
\hspace*{-3.5mm} WP/Wilson & $5.8$ [$0.054(2)$]    & $50$  & $\kappa_1= 0.1535$ & \cite{spqcdr}	\\
\hspace*{-3.5mm} {\color{blue}Set 1}  &    &   $50$   & $\kappa_2= 0.1538$ & 	\\
\hspace*{-3.5mm}   &    &   $50$   & $\kappa_3= 0.1540$ &  	\\
\hspace*{-3.5mm}   &    &   $50$   & $\kappa_4= 0.1541$ &  \\ \hline
{\phantom{\Large{l}}}  {\phantom{\Large{j}}}
\hspace*{-5.5mm} Iwasaki/Clover & $2.1$ [$0.099(0)$]   & $100$  & $\kappa_1= 0.1357$ & \cite{cp-pacs}	\\
\hspace*{-3.5mm}{\color{blue}Set 2}   &    &   $100$   & $\kappa_2= 0.1367$ & 	\\
\hspace*{-3.5mm}   &    &   $100$   & $\kappa_3= 0.1374$ & 	\\
\hspace*{-3.5mm}   &    &   $100$   & $\kappa_4= 0.1382$ & 		\\ \hline
{\phantom{\Large{l}}}   {\phantom{\Large{j}}}
\hspace*{-3.5mm}WP/Clover & $5.29$ [$0.075(1)$]    &   $60$   & $\kappa_1= 0.1355$ & \cite{qcdsf}	\\
\hspace*{-3.5mm} {\color{blue}Set 3}  &    &   $80$   & $\kappa_2= 0.1359$ & 	\\
\hspace*{-3.5mm}   &    &   $100$   & $\kappa_3= 0.1362$ & 	\\  \hline
\end{tabular}\caption{\label{tab:1} \sl Basic information on the sets of unquenched gauge field configurations with $N_f=2$ dynamical Wilson quarks, 
the hopping parameters of which are specified for each set. ``WP" stands for the Wilson-Plaquette gauge action, and Clover is a standard distinction 
to indicate the non-perturbative ${\cal O}(a)$-improved Wilson quark action. More information on each set of configurations can be found in the quoted references. 
All lattice volumes are $24^3\times 48$.}
\end{center}
\vspace*{-3mm}\end{table} 
\subsection{Axial current renormalization constant}
The final ingredient  necessary to relate the results of our calculation to the continuum limit is the appropriate axial current 
renormalization. We prefer to apply the same procedure to all our data sets and determine non-perturbatively the axial renormalization constant .
To avoid any notational ambiguity we stress that in this subsection we discuss only the light bilinear quark non-singlet 
operators,  $P(x) =\bar q(x)\gamma_5 q(x)$, $V_\mu =\bar q(x)\gamma_\mu q(x)$,  $A_\mu =\bar q(x)\gamma_\mu \gamma_5 q(x)$, i.e. 
no reference to the static heavy quark will be needed in this subsection. To evaluate  $Z_A(g_0^2)$ we use the hadronic Ward identity~\cite{ward}, 
which is readily derived by imposing the invariance under  the axial chiral rotations of  $\langle \sum_{\vec x} V_i(x)A_i(0)\rangle$, and $\langle \sum_{\vec x} V_0(x) P(0)\rangle$. 
One then obtains 
\bea\label{WIDS}
&&\hspace*{-4mm}{Z_V^2\over Z_A^2} \langle {\displaystyle \sum_{\vec x}} V_i(x) V_i(0)\rangle =  \langle {\displaystyle \sum_{\vec x}} A_i(x) A_i(0)\rangle\cr 
&&\hspace*{10mm}- Z_V {\displaystyle \int_{\cal V} }d^4z \langle {\displaystyle \sum_{\vec x}} 2 \mawi P(z) V_i(x) A_i(0)\rangle ,   \\
&&\hspace*{-4mm}\langle {\displaystyle \sum_{\vec x}} A_0(x) P(0)\rangle  =\cr
&&\label{WIDS2}\hspace*{12mm}Z_V {\displaystyle \int_{\cal V'} }d^4z \langle {\displaystyle \sum_{\vec x}} 2 \mawi P(z) V_0(x) P(0)\rangle  ,
\eea
where the integration volume ${\cal V}$ (${\cal V}'$) does (does not) include zero.  The bare quark mass defined via the axial Ward identity reads, 
$2 \mawi = \langle \sum_{\vec x} A_0(x) P(0)\rangle/\langle \sum_{\vec x} P(x) P(0)\rangle$.  
For notational simplicity in the above Ward identities we wrote $Z_{V,A}\equiv Z_{V,A}(g_0^2,am_q)$. 

\section{Lattice details and results}

We use the publicly available gauge field configurations generated with $N_f=2$ dynamical light ({\sl ``sea"}) quarks which were produced by using the 
Wilson gauge and Wilson quark actions. 
In table~\ref{tab:1} we provide a basic information on the data sets used in this letter. Concerning the discretized Yang-Mills part,  the configurations explored in this letter were generated by  
the standard Wilson plaquette action and (in one of the sets) by  its improved form, known as the Iwasaki action. The effects of dynamical quarks in the  QCD 
vacuum  fluctuations are simulated by using the Wilson quark action, both the ordinary one, and its non-perturbatively ${\cal O}(a)$-improved version, which is usually referred to as the ``Clover"-action. 
From the publicly available configurations we chose those with  small lattice spacings,  $a\lesssim 0.1$~fm. In  table~\ref{tab:1} we also provide the references containing  
 detailed information about the simulation parameters and the algorithms used in producing these configurations. The values of lattice spacings, given in table~\ref{tab:1}, are 
obtained from $r_0/a$, computed on each of these lattices, extrapolated to the chiral limit and then by choosing $r_0=0.467$~fm.  Other popular choice is $r_0=0.5$~fm,  which would make 
the lattice spacing  $7\%$ larger. To our purpose that error on fixing the lattice spacing is, however, completely immaterial.
We should emphasize that we do not work in the partially quenched situations. Instead, we fix the hopping parameter ($\kappa_q$) of our valence light quark in correlation functions~(\ref{eq:correlators}) and in 
those appearing in eqs.~(\ref{WIDS},\ref{WIDS2}) to be equal to that 
of the corresponding dynamical (``{\it sea}") quark,  also listed in table~\ref{tab:1}.

\begin{figure}
\hspace*{-.7cm}\includegraphics[width=8.7cm,clip]{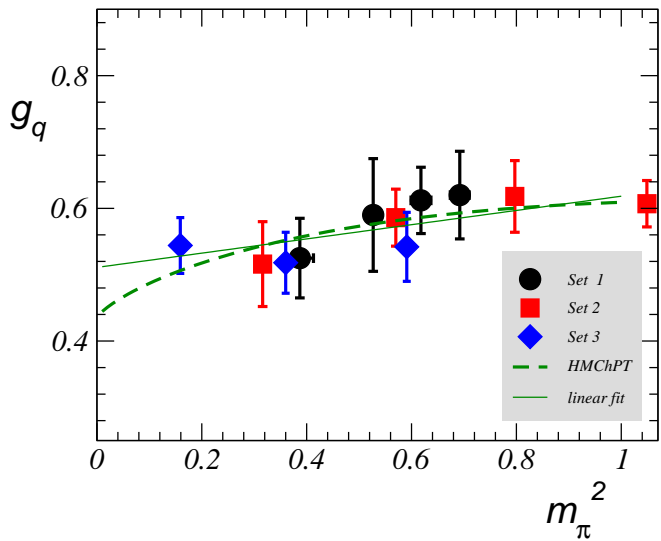}
\caption{\label{plotW} \sl $\hat g_q$ computed from the ratio in eq.~(\ref{gfit}) for all of our lattice data sets listed in table~\ref{tab:1}, after accounting for the axial current renormalization 
constants computed on the same ensembles of gauge field configurations. They are plotted as a function of the light pseudoscalar meson (``pion") mass squared (in $\gev^2$). }
\end{figure}

In this paper we do not use the so-called ``all-to-all" propagators. The feasibility study of using that technique in the computation of $\hat g$-coupling has been made recently in ref.~\cite{onogi}, 
showing the substantial reduction in statistical errors. We plan to adopt that technique in our future studies. 
\begin{table*}[h!!]
\begin{ruledtabular}
\begin{tabular}{|c|c|c|c|c|c|c|} 
{\phantom{\huge{l}}} \raisebox{-.2cm} {\phantom{\huge{j}}}
$\beta$ & $\hspace*{5mm}\kappa_{q}\hspace*{5mm}$ & $\hspace*{4mm}a m_\pi\hspace*{4mm}$ & $\hspace*{4mm}a \mawi \hspace*{4mm}$  & $Z_V(g_0^2,m_q)$  & $Z_A(g_0^2,m_q)$ & $\hspace*{6mm}\hat g_q\hspace*{6mm}$ \\ \hline \hline
{\phantom{\huge{l}}} \raisebox{-.2cm} {\phantom{\huge{j}}}
$5.8$ & $0.1535$ & $0.262(4)$    & $0.0333(6)$  & $0.720(38)$ & $0.908(50)$ & $0.683(52)$	\\
{\phantom{\huge{l}}} \raisebox{-.2cm} {\phantom{\huge{j}}}
{\color{blue}Set 1}           & $0.1538$ & $0.236(4)$    & $0.0260(3)$  & $0.705(41)$ & $0.858(51)$ & $0.714(71)$	\\
{\phantom{\huge{l}}} \raisebox{-.2cm} {\phantom{\huge{j}}}
           & $0.1540$ & $0.221(3)$    & $0.0215(4)$  & $0.643(57)$ & $0.795(70)$ & $0.742(81)$	\\
{\phantom{\huge{l}}} \raisebox{-.2cm} {\phantom{\huge{j}}}
           & $0.1541$ & $0.182(7)$    & $0.0180(4)$  & $0.654(61)$ & $0.827(81)$ & $0.628(62)$	\\
	 \hline \hline
{\phantom{\huge{l}}} \raisebox{-.2cm} {\phantom{\huge{j}}}
$2.1$ & $0.1357$ & $0.631(2)$    & $0.1078(5)$  & $0.742(13)$ & $0.822(13)$ & $0.738(42)$	\\
{\phantom{\huge{l}}} \raisebox{-.2cm} {\phantom{\huge{j}}}
{\color{blue}Set 2}             & $0.1367$ & $0.519(2)$    & $0.0743(3)$  & $0.751(25)$ & $0.839(18)$ & $0.736(56)$	\\
{\phantom{\huge{l}}} \raisebox{-.2cm} {\phantom{\huge{j}}}
           & $0.1374$ & $0.422(2)$    & $0.0513(4)$  & $0.754(38)$ & $0.847(26)$ & $0.691(55)$	\\
{\phantom{\huge{l}}} \raisebox{-.2cm} {\phantom{\huge{j}}}
           & $0.1382$ & $0.298(2)$    & $0.0252(4)$  & $0.752(82)$ & $0.829(50)$ & $0.622(61)$	\\
	 \hline \hline
{\phantom{\huge{l}}} \raisebox{-.2cm} {\phantom{\huge{j}}}
$5.29$ & $0.1355$ & $0.327(2)$    & $0.0357(3)$  & $0.772(2)$ & $0.793(5)$ & $0.683(63)$	\\
{\phantom{\huge{l}}} \raisebox{-.2cm} {\phantom{\huge{j}}}
  {\color{blue}Set 3}           & $0.1359$ & $0.245(2)$    & $0.0206(2)$  & $0.770(4)$ & $0.786(9)$ & $0.658(61)$	\\
{\phantom{\huge{l}}} \raisebox{-.2cm} {\phantom{\huge{j}}}
           & $0.1362$ & $0.155(2)$    & $0.0086(3)$  & $0.760(10)$ & $0.779(14)$ & $0.711(62)$	\\
\end{tabular}\caption{\label{tab:res}\footnotesize Direct numerical results extracted from the correlation functions calculated on all of the ensembles of the lattices 
with parameters enumerated in table~\ref{tab:1}. }
\end{ruledtabular}
\end{table*}
In table~\ref{tab:res} we provide the list of all results relevant to the subject of this letter, that we directly extracted from the correlation functions  computed on all lattices from table~\ref{tab:1}. 
For an easier comparison, the values of the pseudoscalar light meson masses, as well as of the bare light quark masses inferred from the axial vector Ward identity, are given in lattice units.  
They are fully consistent with those reported in refs.~\cite{spqcdr,cp-pacs,qcdsf}. Concerning the renormalization constants 
$Z_{V,A}$, they are obtained from the {\sl ``light-light"} correlation functions which we computed on the lattice and then combined to verify the Ward identities in eqs.~(\ref{WIDS},\ref{WIDS2}). After inspection, 
we found the common plateau-region for all our $11$ data sets to be between $10\leq t\leq 14$.  Finally, in the three-point correlation function in eq.~(\ref{eq:correlators}) the fixed source operator is set at $t_y=13$, 
and --as already mentioned-- the results for $\hat g_q$ are obtained from the fit to a constant in eq.~(\ref{gfit}),  on the interval $5\leq t_x\leq 8$.  We checked that our results remain stable when $t_y=12$.  
Directly extracted values for the bare couplings $\hat g_q$,  from all of the lattice data sets considered in this letter, i.e. before multiplying them by its corresponding $Z_A$,  are listed in table~\ref{tab:res}. 
In fig.~\ref{plotW}, instead, we plot the renormalized coupling $\hat g_q$, as a function of the squared light-light pseudoscalar meson (``pion"), mass now given in physical units (in $\gev^2$).
That conversion is made by computing $r_0 m_\pi$ for each of our data sets and then use $r_0=0.467$~fm (or, $r_0=2.367~\gev^{-1}$).  We reiterate that opting for 
$r_0=0.5$~fm  ($r_0=2.534~\gev^{-1}$), does not alter our final results in any significant way. 

The last step to reach the coupling $\hat g$, which is our final goal, is to make the extrapolation to the chiral limit. To that end we attempt either a simple linear fit or a fit guided by the expression derived in HMChPT~\cite{hmg}, i.e.,
\bea
&&\label{lin}\hat g^q = \hat g_{\rm lin}\left( 1 + c_{\rm lin}  m_\pi^2\right)\,,   \\
&&\label{chi}\hat g^q = \hat g_0 \left[ 1 - {4\hat g_0^2 \over (4 \pi f)^2} m_\pi^2 \log(m_\pi^2) + c_{0} m_\pi^2\right]\,,
\eea
where $\hat g_0$ is then the soft pion coupling that is to be used in applying the HMChPT formulas when extrapolating the 
phenomenologically interesting quantities computed on the lattice to the physical light quark mass limit.
From  fig.~\ref{plotW} it is obvious that this task is quite difficult if one is doing it separately for each $\beta$. 
More specifically, applying the linear fit~(\ref{lin}) to each of our data sets we obtain 
\bea
 \hat g_{\rm lin} = \{ (0.40\pm 0.15)_1, (0.52\pm 0.07)_2, (0.54\pm 0.06)_3\} \,,
\eea
while from the fit to HMChPT~(\ref{chi}) we get
\bea
 \hat g_{0} = \{ (0.36\pm 0.11)_1, (0.43\pm 0.04)_2, (0.48\pm 0.04)_3\} \,,
\eea
where the index on the right hand side labels the data sets like in the tables~\ref{tab:1} and \ref{tab:res}. As it could have been anticipated from eq.~(\ref{chi}), the results of the HMChPT fit ($\hat g_0$) are lower 
than the results of linear extrapolation ($ \hat g_{\rm lin}$). The values obtained from different sets are consistent within the errors. It is obvious that we cannot make a precision determination of this coupling yet, 
but it is clear that the unquenched lattice data also point to the fact that the $\hat g$ coupling  is considerably smaller in the static-heavy quark limit than 
in the case of the heavy charm quark. 
Since the heavy quark is only a spectator, this information --that the $1/m_h^n$-corrections are large-- is significant, and somewhat surprising. If one simply feeds the difference by a linear $1/m_c$-term, it is quite interesting 
to notice that from the light cone QCD sum rules one get a similar size is such a correction in spite of the fact that the absolute value for the pionic couplings were considerably underestimated~\cite{KRWY}.

Since we are not aiming at a percent-level precision determination of this coupling we can try and see what happens if all the data are combined and fit them together to eqs.~(\ref{lin},\ref{chi}). We are, of course, aware 
that our three sets suffer from different discretization errors but since the lattice spacing is small ($a < 0.1$~fm) and the common renormalization procedure has been applied to all of them, it is reasonable to assume 
that the remaining discretization errors are not likely to matter, in view of our statistical error  ($\sim 10\%$). 
If we combine all of our data, we then obtain 
\bea
 \hat g_{\rm lin} = 0.51\pm 0.04\,,\quad c_{\rm lin}=(0.21\pm 0.12)~\gev^{-1}\,,
\eea
while with the HMChPT formula~(\ref{lin}) we have
\bea
 \hat g_{0} = 0.44\pm 0.03\,,\quad c_0=(0.40\pm 0.12)~\gev^{-1}\,.
\eea
Another possibility is to exclude the data with $m_\pi^2\geq 0.6~\gev^2$, which gives $ \hat g_{0} = 0.46\pm 0.04$. 
We also checked that our resulting $\hat g_0$ is insensitive to the variation of  $f \in (120, 132)$~MeV, latter being $f_\pi^{\rm phys.}$.

Before concluding we should compare our result to the existing unquenched value for $\hat g$ reported in ref.~\cite{onogi}. 
The main advantage of the calculation presented in ref.~\cite{onogi} with respect to ours is that they used the so-called all-to-all light quark propagators 
so that their resulting  statistical errors are much smaller.  However the lattices we used here are finer and the associated discretization errors should be smaller. 
In addition, here we also use various gauge and quark actions, to show that our results are robust in that respect too (of course within our error bars). 
A reasonable comparison with ref.~\cite{onogi} can be made by using our results from Set-2 because these data correspond to the same gauge and quark actions 
as those used in ref.~\cite{onogi}, although the lattice spacing we use here is smaller.  Comparing the bare quantities, we see that --for example-- when the pion mass is $m_\pi\approx 0.75$~GeV, 
from ref.~\cite{onogi} we read $g_q^{\beta=1.8}=0.68(1)$, $g_q^{\beta=1.95}=0.69(1)$,  while our $g_q^{\beta=2.1}=0.69(6)$. Therefore they fully agree although our statistical errors
are much larger. Using the perturbative (boosted) 1-loop expression (bpt), the result for the overall renormalization constant in all three cases are equal among themselves within less 
than $1\%$, so that the renormalization constants computed in bpt would not spoil this comparison, which seems to indicate that the discretization errors are indeed small. 
The step in which we go beyond ref.~\cite{onogi} is that we evaluate the axial renormalization constant non-perturbatively, $Z_A^{\rm npr}$.  This is particularly important 
when using the data obtained with Iwasaki gauge action because  in that case the strong coupling is very large and the use of perturbation theory is far from being justified. 
Various boosting procedures can lead to various estimates of $Z_A$. We show that the boosting procedure used in refs.~\cite{onogi,cp-pacs} leads to the values very close to our 
non-perturbative estimate. More precisely, at our $\beta=2.1$, we have $Z_A^{\rm npr}/Z_A^{\rm bpt} =0.90(1),0.94(2),0.97(3),0.97(6)$, when going from the heaviest to the lightest quark mass.

\section{\label{conclusions}Conclusions}
In this letter we report on the results of our calculations of the soft pion coupling to the lowest lying doublet of static heavy-light mesons. From our computations, in which we use the fully unquenched set-up  and 
three different sets of gauge field configurations, all produced with Wilson gauge and fermion actions, we obtain that $ \hat g_{0} =0.44\pm 0.03{}^{+0.07}_{-0.00}$. The second error reflects the uncertainty due to chiral 
extrapolation and it is the difference between the results of linear fit and the fit in which HMChPT is used. If our result is to be used in the chiral extrapolations of the phenomenologically relevant quantities in $B$-physics the second error should be considered as flat. The reason is that our central value is obtained via HMChPT fit, but since the domain of applicability of HMChPT is still unclear~\cite{su2} --as of now-- both results (extrapolated 
linearly or by using the chiral loop correction) are equally valid. 

On the more qualitative level, our results  show/confirm that this coupling is smaller in the static limit than what is obtained when the heavy quark is propagating and is of the mass equal to that of the physical charm quark, $ \hat g_{\rm charm} =0.68\pm 0.07$~\cite{gDDpi-unq}.  It is intriguing that the ${\cal O}(1/m_c^n)$ corrections 
are quite large for the quantity in which the heavy quark contributes only as a spectator. That feature can be safely studied on the lattice by means of  the relativistic heavy quark action of ref.~\cite{RQA}, and we plan  to do  
such a study. An obvious perspective concerning the determination of  $ \hat g_{0}$ is to further reduce the errors, both statistical (by using the ``all-to-all" propagator technique, like in ref.~\cite{onogi}), and the systematic ones 
(in particular those associated with  chiral extrapolations). Once a per-cent accuracy is reached, it will be important to study carefully the effects of mixing with the lowest heavy-light excited states [those with $j_\ell^P=(1/2)^+$]. 
The expression  derived in HMChPT which accounts for those effects in $\hat g_q$ already exist (first paper in ref.~\cite{hmg}), but their use requires the knowledge of another pionic coupling, the one that parametrizes the $S$-wave 
pion emitted in the transition from a   $(1/2)^+ \to (1/2)^-$ states. Finally, the numerical tests concerning the impact of inclusion of heavier quarks in the vacuum fluctuations ($s$ and $c$) on the size of $\hat g_0$, would be 
highly welcome too. 


\begin{acknowledgments}
We thank  the SPQcdR, CP-PACS and QCDSF collaborations for making their gauge field configurations publicly available, 
The supports of  `Flavianet'  (EU contract MTRN-CT-2006-035482) and of the ANR (contract ÒDIAMÓ ANR-07-JCJC-0031) are kindly acknowledged too.  
Laboratoire de Physique Th\'eorique is unit\'e mixte de Recherche du CNRS - UMR 8627.
\end{acknowledgments}



\begin{thebibliography}{99}


\bibitem{Casalbuoni}
R.~Casalbuoni, A.~Deandrea, N.~Di Bartolomeo, R.~Gatto, F.~Feruglio and G.~Nardulli,
Phys.\ Rept.\  {\bf 281} (1997) 145
[hep-ph/9605342].

\bibitem{su2}
  D.~Becirevic, S.~Fajfer and J.~F.~Kamenik,
  JHEP {\bf 0706} (2007) 003
  [arXiv:hep-ph/0612224];


\bibitem{CLEO}
A.~Anastassov {\it et al.}  [CLEO Collaboration],
Phys.\ Rev.\ D {\bf 65} (2002) 032003
[hep-ex/0108043].


\bibitem{KRWY}
A.~Khodjamirian {\it et al.}, 
Phys.\ Lett.\ B {\bf 457} (1999) 245
[hep-ph/9903421],
see also P.~Ball and R.~Zwicky
Phys.\ Rev.\ D {\bf 71} (2005) 014015
[hep-ph/0406232].


\bibitem{dirac}
  D.~Becirevic and A.~L.~Yaouanc,
  JHEP {\bf 9903} (1999) 021
  [arXiv:hep-ph/9901431].


\bibitem{gDDpi}
A.~Abada {\it et al.},
Phys.\ Rev.\ D {\bf 66} (2002) 074504
[hep-ph/0206237].


\bibitem{gDDpi-unq}
  D.~Becirevic and B.~Haas,
  arXiv:0903.2407 [hep-lat].



\bibitem{us9}
  D.~Becirevic, J.~Charles, A.~LeYaouanc, L.~Oliver, O.~Pene and J.~C.~Raynal,
  JHEP {\bf 0301} (2003) 009
  [arXiv:hep-ph/0212177].


\bibitem{g-ukqcd-orsay}
G.~M.~de Divitiis {\it et al.} [UKQCD Collaboration],
JHEP {\bf 9810} (1998) 010
[hep-lat/9807032]; 
A.~Abada {\it et al.},
JHEP {\bf 0402} (2004) 016
[hep-lat/0310050].


\bibitem{onogi}
  H.~Ohki, H.~Matsufuru and T.~Onogi,
  Phys.\ Rev.\  D {\bf 77} (2008) 094509
  [arXiv:0802.1563 [hep-lat]].



\bibitem{Eichten-Hill}
E.~Eichten and B.~Hill,
Phys.\ Lett.\ B {\bf 234} (1990) 511.



\bibitem{knechtli}
  A.~Hasenfratz and F.~Knechtli,
  Phys.\ Rev.\  D {\bf 64} (2001) 034504
  [arXiv:hep-lat/0103029].


\bibitem{alpha}
  M.~Della Morte, S.~Durr, J.~Heitger, H.~Molke, J.~Rolf, A.~Shindler and R.~Sommer
                  [ALPHA Collaboration],
  Phys.\ Lett.\  B {\bf 581} (2004) 93
  [Erratum-ibid.\  B {\bf 612} (2005) 313]
  [arXiv:hep-lat/0307021].
  

  
\bibitem{Boyle}
P.~Boyle  [UKQCD Collaboration],
J.\ Comput.\ Phys.\  {\bf 179} (2002) 349
[hep-lat/9903033].

\bibitem{ward}
G.~Martinelli, S.~Petrarca, C.~T.~Sachrajda and A.~Vladikas,
  Phys.\ Lett.\  B {\bf 311} (1993) 241
  [Erratum-ibid.\  B {\bf 317} (1993) 660];
  M.~Crisafulli, V.~Lubicz and A.~Vladikas,
  Eur.\ Phys.\ J.\  C {\bf 4} (1998) 145
  [arXiv:hep-lat/9707025].
T.~Bhattacharya, R.~Gupta, W.~J.~Lee and S.~R.~Sharpe,
  Phys.\ Rev.\  D {\bf 63} (2001) 074505
  [arXiv:hep-lat/0009038].

\bibitem{emmanuel}
  D.~Becirevic, E.~Chang and A.~Le~Yaouanc, 
  arXiv:0905.3352 [hep-lat].

\bibitem{spqcdr}
  D.~Becirevic {\it et al.},
  Nucl.\ Phys.\  B {\bf 734} (2006) 138
  [arXiv:hep-lat/0510014].

\bibitem{cp-pacs}
  A.~Ali Khan {\it et al.}  [CP-PACS Collaboration],
  Phys.\ Rev.\  D {\bf 65} (2002) 054505
  [Erratum-ibid.\  D {\bf 67} (2003) 059901]
  [arXiv:hep-lat/0105015].

\bibitem{qcdsf}
A.~A.~Khan {\it et al.},
  Phys.\ Rev.\  D {\bf 74} (2006) 094508
  [arXiv:hep-lat/0603028];
M.~Gockeler {\it et al.},
  Phys.\ Rev.\  D {\bf 73} (2006) 054508
  [arXiv:hep-lat/0601004].



\bibitem{hmg}
S.~Fajfer and J.~F.~Kamenik,
  Phys.\ Rev.\  D {\bf 74} (2006) 074023
  [arXiv:hep-ph/0606278]; 
  D.~Becirevic, S.~Prelovsek and J.~Zupan,
  Phys.\ Rev.\  D {\bf 67} (2003) 054010
  [arXiv:hep-lat/0210048].
I.~W.~Stewart,
Nucl.\ Phys.\ B {\bf 529}, 62 (1998).


\bibitem{RQA}
  A.~X.~El-Khadra, A.~S.~Kronfeld and P.~B.~Mackenzie,
  Phys.\ Rev.\  D {\bf 55} (1997) 3933
  [arXiv:hep-lat/9604004];
  N.~H.~Christ, M.~Li and H.~W.~Lin,
  Phys.\ Rev.\  D {\bf 76} (2007) 074505
  [arXiv:hep-lat/0608006];
Y.~Kayaba {\it et al.}  [CP-PACS Collaboration],
  JHEP {\bf 0702} (2007) 019
  [arXiv:hep-lat/0611033].


\end{thebibliography}

\end{document}